\documentclass[prb,amsmath,amssymb,superscriptaddress,twocolumn]{revtex4}
\usepackage{float}
\usepackage{amsmath}
\usepackage{amssymb}
\usepackage{amsfonts}
\usepackage{euscript}
\usepackage{enumerate}
\usepackage{hhline}
\usepackage{pslatex}
\usepackage{tabularx}
\usepackage[usenames,dvipsnames]{xcolor}

\usepackage{graphicx}

\usepackage{dcolumn}
\usepackage{bm}

\usepackage{lipsum}

\makeatletter
\renewcommand{\@biblabel}[1]{#1. }
\renewcommand{\@dotsep}{500}
\renewcommand{\@pnumwidth}{0em}
\renewcommand{\l@figure}[2]{
\@dottedtocline{1}{1.5em}{2em}{Figure #1}
\vspace{15pt}}

\newcommand{\xl}[1]{\textcolor{black}{#1}}

\usepackage[normalem]{ulem}

\begin{document}
\title{Direct growth of crack-free stoichiometric silicon nitride on sapphire for wide-band photonics}

\author{Zhaohui Ma}
\affiliation{Microsystems and Nanotechnology Division, Physical Measurement Laboratory, National Institute of Standards and Technology, Gaithersburg, MD 20899, USA}
\affiliation{Joint Quantum Institute, University of Maryland College Park, College Park, MD 20742, USA}
\author{Ashish Chanana}
\affiliation{American Institute of Manufacturing Integrated Photonics, Research Foundation SUNY, Albany, New York, USA}
\author{Zhiquan Yuan} 
\affiliation{Joint Quantum Institute, University of Maryland College Park, College Park, MD 20742, USA}
\author{Khoi Hoang}
\affiliation{Joint Quantum Institute, University of Maryland College Park, College Park, MD 20742, USA}
\author{Yiliang Bao} 
\affiliation{Microsystems and Nanotechnology Division, Physical Measurement Laboratory, National Institute of Standards and Technology, Gaithersburg, MD 20899, USA}
\author{Daron Westly}
\affiliation{Microsystems and Nanotechnology Division, Physical Measurement Laboratory, National Institute of Standards and Technology, Gaithersburg, MD 20899, USA}
\author{Kartik Srinivasan} \email{kartik.srinivasan@nist.gov}
\affiliation{Microsystems and Nanotechnology Division, Physical Measurement Laboratory, National Institute of Standards and Technology, Gaithersburg, MD 20899, USA}
\affiliation{Joint Quantum Institute, University of Maryland College Park, College Park, MD 20742, USA}
\author{Xiyuan Lu}\email{xnl9@umd.edu}
\affiliation{Microsystems and Nanotechnology Division, Physical Measurement Laboratory, National Institute of Standards and Technology, Gaithersburg, MD 20899, USA}
\affiliation{Joint Quantum Institute, University of Maryland College Park, College Park, MD 20742, USA}

\date{\today}

\begin{abstract}
\noindent Low-loss stoichiometric silicon nitride(Si$_3$N$_4$) on oxidized silicon has emerged as the major wide-band photonics platform for nonlinear and quantum photonics. These films, deposited by low-pressure chemical vapor deposition(LPCVD), are well-known to have around 1000~MPa of residual tensile stress that often leads to cracking issues for film thickness above 400~nm. Though various stress mitigation approaches can isolate or prevent the cracking, this stress issue remains a major challenge in fabrication. On the other hand, low-residual-stress Si$_3$N$_4$ on sapphire has been reported, and recently explored for photonics at telecommunication wavelengths. Here, we advance its development through systematic characterization consisting of thin-film stress analysis, ellipsometry and prism coupling measurements of refractive index and loss, and swept-wavelength spectroscopy of fabricated photonic devices across a broad range of wavelengths. We demonstrate direct growth of crack-free Si$_3$N$_4$ up to 2~$\mu$m in thickness with only 200~MPa stress. Our fabricated devices show a propagation loss of 0.35~dB/cm around 780~nm, 0.37~dB/cm around 1064~nm, and 0.25~dB/cm around 1630~nm. In addition, prism coupling measurements indicate film losses below 1~dB/cm at wavelengths down to 500~nm. Our work highlights Si$_3$N$_4$ on sapphire as a promising wide-band photonics platform.
\end{abstract}  

\maketitle

{
  \let\thefootnote\relax 
  \footnotetext{\hspace{-0.1in}$^{\ddagger}$~This document is preliminary and is intended for peer review conducted by a journal.}
}

\section{Introduction}
\begin{figure*}[t!]
\centering\includegraphics[width=0.96\linewidth]{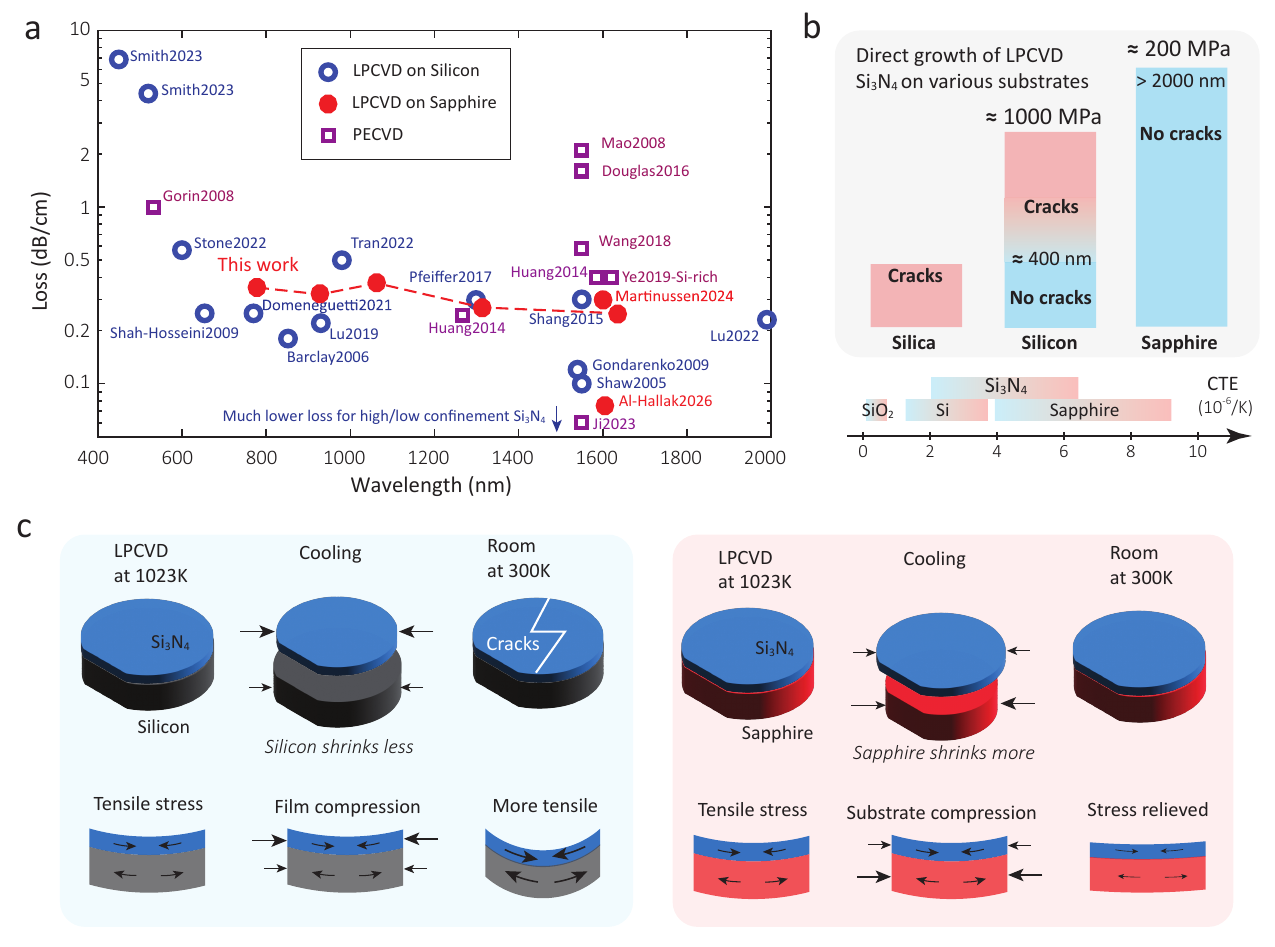}
\caption{\textbf{Comparison of low-pressure chemical vapor deposition (LPCVD) of stoichiometric silicon nitride (Si$_3$N$_4$) on different substrates.} \textbf{a,} Loss in silicon nitride photonic integrated circuits (PICs) as a function of wavelength, including the most commonly utilized LPCVD Si$_3$N$_4$ (blue circles)~\cite{Smith_OptMaterExpress_2023, Stone_APLPhoton_2022, Shah_Hosseini_OptExpress_2009, Domeneguetti_Optica_2021, Barclay_ApplPhysLett_2006, Lu_Optica_2019, Tran_Nature_2022, Pfeiffer_Optica_2017, Shaw_SPIE_2005, Gondarenko_OptExpress_2009, Shang_OptExpress_2015, Lu_OptLett_2022} and plasma-enhanced chemical vapor deposition (PECVD) (red squares)~\cite{Gorin_OptExpress_2008, Huang_OptExpress_2014, Ji_LaserPhotonRev_2023, Ye_OptLett_2019, Wang_OptExpress_2018, Douglas_OptMaterExpress_2016,Mao_OptExpress_2008} films. Results for LPCVD Si$_3$N$_4$ on sapphire wafers (including this work and two others \cite{Martinussen_OptExpress_2024, Al-Hallak_arXiv_2026}) are in red filled circles. \textbf{b,} 
LPCVD Si$_3$N$_4$ film growth is often limited by tensile stress of around 1000~MPa when grown on silicon substrates, resulting in cracks above a certain thickness (center). For silica substrate with smaller coefficient of thermal expansion (CTE) than Si$_3$N$_4$, growth of crack-free Si$_3$N$_4$ film has not been reported (left), while on sapphire substrates with a larger CTE than Si$_3$N$_4$ (right), we report crack-free film growth up to 2~$\mu$m thickness. \textbf{c,} The cracks in Si$_3$N$_4$ films grown on silicon wafers typically happen not during the LPCVD growth step that is performed at elevated temperature (1023~K to 1123~K), but instead during or after the cooling process to room temperature. The major reason is that the smaller CTE of Si (though relatively close to that of Si$_3$N$_4$) creates extra compression relative to Si$_3$N$_4$ that leads to more residual tensile stress (left). In contrast, a sapphire substrate can relieve the residual tensile stress (right) because of its larger CTE than Si$_3$N$_4$.}
\label{Fig1}
\end{figure*}

\vspace{-0.15in}
\noindent Silicon nitride on insulator has been the most extensively studied wide-band photonic integrated circuit (PIC) platform to date~\cite{Moss_NatPhoton_2013,blumenthal_silicon_2018,poonSiliconPhotonicsVisible2024}, with photonic waveguides and resonators fabricated across a wide range of wavelengths across visible, near-infrared, and infrared regions (Fig.~\ref{Fig1}(a)). Silicon nitride has a wide transparency window reported from 400~nm to 6.7~$\mu$m, with somewhat elevated loss ($>$~2~dB/cm) at deep visible wavelengths~\cite{Smith_OptMaterExpress_2023}, improved performance in the near-infrared ($\approx$~0.2~dB/cm), particularly at telecommunications wavelengths that are the focus of concerted research efforts ($\approx$~0.01~dB/cm)~\cite{Ji_APLPhoton_2021}, as well as low-loss performance ($<$~0.1~dB/cm) in the $3$~$\mu$m to 5~$\mu$m mid-infrared range, where the major limitation is the silicon dioxide cladding layers~\cite{Miller_Optica_2017}. This low-loss PIC platform encompasses ultra-high quality factor ($Q$) cavities using thin~\cite{Spencer_Optica_2014,Puckett_NatCommun_2021, Li_Optica_2023} or thick~\cite{Pfeiffer_Optica_2016,Ji_Optica_2017,Ji_APLPhoton_2021} waveguiding layers for enhancing light-matter interactions and through heterogeneous integration, can be combined with active and electro-optical materials~\cite{Xiang_Optica_2020, Churaev_NatCommun_2023}. Thick silicon nitride films are especially relevant for nonlinear photonics~\cite{dutt_nonlinear_2024}, enabling processes such as Kerr frequency comb generation~\cite{Gaeta_NatPhoton_2019, Chang_NatPhoton_2022}, optical parametric oscillation~\cite{Lu_PIC_OPOs_Optica_2026}, and photo-induced second harmonic generation~\cite{Lu_NatPhoton_2021,Nitiss_NatPhoton_2021}.

Many growth methods are used for silicon nitride, with low-pressure chemical vapor deposition (LPCVD)~\cite{Smith_OptMaterExpress_2023, Stone_APLPhoton_2022, Shah_Hosseini_OptExpress_2009, Domeneguetti_Optica_2021, Barclay_ApplPhysLett_2006, Lu_Optica_2019, Tran_Nature_2022, Pfeiffer_Optica_2017, Shaw_SPIE_2005, Gondarenko_OptExpress_2009, Shang_OptExpress_2015, Lu_OptLett_2022} and plasma-enhanced chemical vapor deposition (PECVD)~\cite{Gorin_OptExpress_2008, Mao_OptExpress_2008, Huang_OptExpress_2014, Douglas_OptMaterExpress_2016, Wang_OptExpress_2018, Ye_OptLett_2019, Ji_LaserPhotonRev_2023,bose_anneal-free_2024} being widely used for photonics, while other techniques such as radio-frequency magnetron sputtering~\cite{Signore_OptMater_2012}, reactive sputtering~\cite{friggOpticalFrequencyComb2020,zhang_lowtemperature_2024}, atomic-pressure chemical vapor deposition~\cite{Perez_SCT_2005}, and molecular beam epitaxy (MBE)~\cite{Katzer_JVSTB_2014} have also been investigated. Among these methods, LPCVD growth of stochiometric silicon nitride (Si$_3$N$_4$) has been the major approach to date, particularly for nonlinear photonics, due to its low losses and repeatable control of thickness and refractive index dispersion, as well as large wafer capacity per furnace run.

However, there is a longstanding issue of excessive film stress in LPCVD growth of Si$_3$N$_4$ on oxidized silicon wafers~\cite{Pan_JES_1985}. This process results in a large residual tensile stress around 1000~MPa, which tends to form cracks across the wafer for film thickness above 400~nm. These thicker films are often needed for the aforementioned nonlinear photonic applications. This stress issue is a major limitation to Si$_3$N$_4$ fabrication, rendering dicing almost impossible, chemical processing quite limited, and wafer spinning needing additional care. As a result, many stress mitigation approaches have been demonstrated. The approaches include the use of deep trenches that serve as crack barriers before~\cite{Luke_OptExpress_2013,xuan_high-q_2016}, in between~\cite{wu_stress-released_2020,liu_fabrication_2025}, and after film growth, thermal cycling and wafer rotation when splitting the film growth into multiple steps~\cite{Dirani_ApplPhysLett_2018}, and damascene processing~\cite{Pfeiffer_Optica_2016}, with thickness up to 2.3~$\mu$m demonstrated~\cite{Guo_NatPhoton_2018}. However, all these approaches only relieve the stress in certain critical areas, but they do not reduce the stress in the unpatterned wafer. They all require additional processing steps during the wafer growth, increasing fabrication complexity, and posing risks of reducing device yield.

On the other hand, low-residual-stress LPCVD Si$_3$N$_4$ on sapphire has been reported~\cite{Habermehl_JVSTA_2018}, and recently explored in photonics at telecommunications wavelengths~\cite{Martinussen_OptExpress_2024, Al-Hallak_arXiv_2026}, offering a promising solution to the longstanding stress challenge. However, broad adoption of this platform and its natural variants (such as Si$_3$N$_4$ on SiO$_2$ on sapphire) requires systematic film stress analysis, photonic device fabrication, and optical characterization. In this paper, we provide further validation of the LPCVD Si$_3$N$_4$ on sapphire platform across all these fronts. First, we provide thorough stress analysis confirming that low residual tensile stress can be achieved by using sapphire as a substrate instead of silicon, due to sapphire's higher coefficient of thermal expansion, with crack-free films as thick as 2~$\mu$m grown. Next, we perform ellipsometry and prism coupling measurements of refractive index and loss in deposited thin films at wavelengths between 400~nm and 1650~nm, providing essential information for photonic device development. To that end, we fabricate microring resonators and perform swept-wavelength spectroscopy across multiple wavelength bands between 780~nm and 1650~nm, with low-loss ($<$~0.4~dB/cm) shown across all bands. Our work demonstrates that LPCVD Si$_3$N$_4$ on sapphire is a promising platform for wide-band linear and nonlinear photonic integrated circuits without stress mitigation.

\begin{figure*}[t!]
\centering\includegraphics[width=0.96\linewidth]{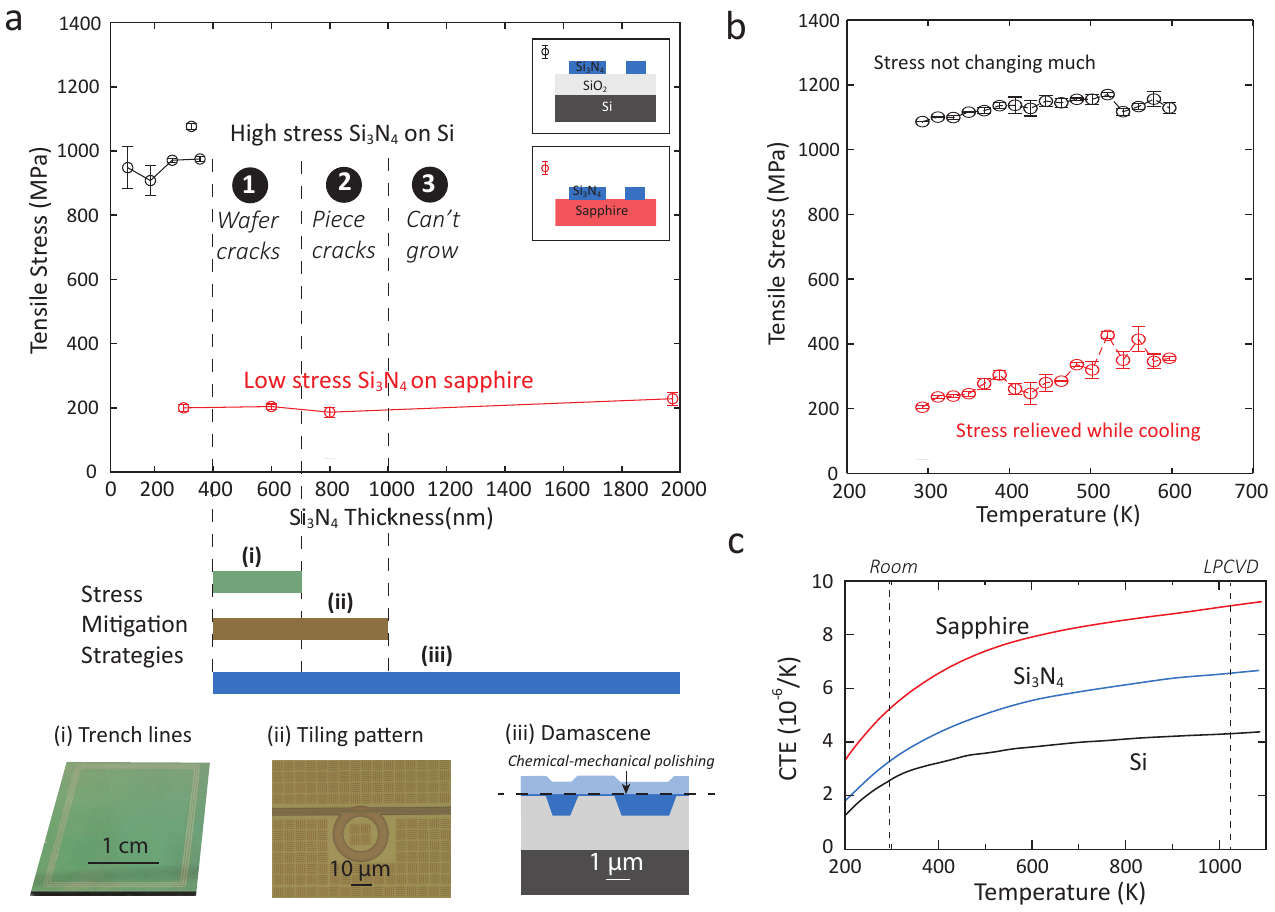}
\caption{\textbf{Stress characterization of LPCVD Si$_3$N$_4$ on a sapphire substrate (SiNoSa) compared to Si$_3$N$_4$ on a silicon substrate (SiNoSi)}. \textbf{a,} SiNoSa has much lower stress (red data points) and can be grown crack-free without substrate modification for film thicknesses up to at least 2~$\mu$m. In contrast, SiNoSi (black data points) cracks easily for films between 350~nm to 400~nm thickness across a wafer, at thickness $>$600~nm for a piece, and is very challenging to grow at all (without substrate modification) at  thickness $>$1000~nm. These two substrate configurations we consider are shown to the right. In the case of SiNoSi, \xl{three stress mitigation} approaches that have been used for thick Si$_3$N$_4$ films are shown at the bottom. (i) we have found that trench lines work for film thickness from 400~nm to 700~nm. (ii) Tiling patterns that incorporate features between 10~$\mu$ to 100~$\mu$m to prevent accumulation of film stress in otherwise unpatterned areas enable film thickness up to 1000~nm and better reproducibility across full wafers. (iii) Damascene processing allows even thicker films to be grown, e.g., at levels up to 2~$\mu$m. It is typically followed by chemical mechanical polishing, and is often used together with tiling pattern for crack termination. \textbf{b,} Temperature-dependent stress measurement for SiNoSa (red) and SiNoSi (black). The uncertainty intervals are \xl{one-standard deviation value} from measurements of stress at six rotational angles (30~$^{\circ}$ steps). \textbf{c,} Literature values for the temperature-dependent CTE of sapphire (red)~\cite{White_IJT_1997}, Si$_3$N$_4$ (solid blue)~\cite{Fang_JAP_2003}, and silicon (black)~\cite{Watanabe_IJT_2004}.}
\label{Fig2}
\end{figure*}


\begin{figure*}[t!]
\centering\includegraphics[width=0.96\linewidth]{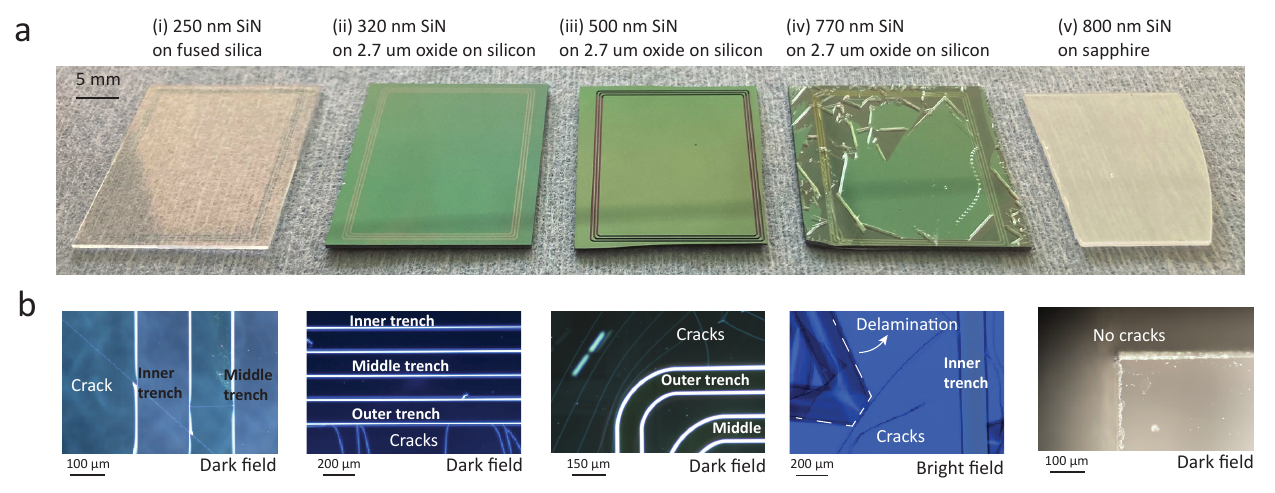}
\caption{\textbf{Cracking analysis of LPCVD Si$_3$N$_4$ grown on different substrates.} \textbf{a,} Five chips on a cleanroom paper, with different thickness of LPCVD Si$_3$N$_4$ on fused silica (i), oxidized silicon (ii-iv), and sapphire (v). The furnace growth parameters are held fixed for all samples. The area that is interior to the deep trenches in samples (i-iv) is around 1.6~cm x 2.1~cm. Sample (v) does not have deep trenches. \textbf{b,} Fused silica, silicon, and sapphire substrates exhibit different cracking phenomena. Bright-field and dark-field microscope images are used to inspect the pieces for cracks. (i) Cracking is most severe for LPCVD Si$_3$N$_4$ on fused silica wafers. (ii,iii) For oxidized silicon wafers, crack propagation can often stop at the outside trench. (iv) For thicker films on oxidized silicon wafers, the film has cracks directly after deposition. Moreover, it gets much worse after a few hours, with more cracks and severe film delamination. (v) On sapphire wafers, no cracks even near the diced edges, so no trench lines are needed. In (i-iv) the trenches are formed by three boxes with each trench 160 micrometers wide with oxide fully removed.}
\label{Fig3}
\end{figure*}

\vspace{-0.15in}
\section{Stress analysis}
\vspace{-0.15in}
\noindent Although LPCVD Si$_3$N$_4$ growth is primarily reported on silicon, we first take a step back and survey its growth on multiple substrate options, including fused silica, silicon, and sapphire, as illustrated in Fig.~\ref{Fig1}(b). Empirically, we find that while silicon has the closest coefficient of thermal expansion (CTE) to Si$_3$N$_4$, films grown on it are considerably worse than those grown on sapphire in terms of stress and cracks. Another important but less studied reference point is a fused silica substrate, where we found it is challenging to grow even 100~nm Si$_3$N$_4$ thick films without cracking, even when employing stress-mitigation approaches such as substrate trench lines. Examination of these three substrates and their outcomes supports the claim that Si$_3$N$_4$ favors the substrate with larger CTE, which we will further substantiate in Section II and III. The principle idea behind this claim is illustrated in Fig.~\ref{Fig1}(c), where substrate compression during cooling reduces the residual tensile stress of the film. For Si$_3$N$_4$ growth on a silicon substrate (Fig.~\ref{Fig1}(c), left), when the LPCVD film is deposited, there is internal tensile stress at the (elevated) growth temperature (typically 1023~K to 1123~K). During the cooling process after film growth, silicon's smaller CTE than Si$_3$N$_4$ causes it to shrink less. This compresses the Si$_3$N$_4$ film and make the residual stress more tensile, leading to more bowing of the wafer and cracks in films thicker than 350~nm. Even when the film does not crack immediately after growth, any external stress, like handling by metal tweezers, chemical processes like RCA cleaning, and mechanical process like the spinning, rinsing, and drying (SRD) can crack or even shatter the film. The above behavior is supported by the observation that cracking almost always happens after the growth or after cooling, but not during the growth step itself. On the other hand, sapphire is very good at hosting LPCVD Si$_3$N$_4$ growth (Fig.~\ref{Fig1}(c), right). At elevated temperature, Si$_3$N$_4$ is still deposited with tensile stress, but during the cooling process, because sapphire shrinks more than silicon does, the overall residual stress is relieved and much less tensile. Thus, LPCVD Si$_3$N$_4$ grown on sapphire is free from cracks, and can go through chemical and mechanical processes like any low-stress Si-rich silicon nitride on silicon, though being stoichiometric Si$_3$N$_4$.

In Fig.~\ref{Fig2}, we show stress measurements that support the stress relief hypothesis described above and in Fig.~\ref{Fig1}(b). We compare Si$_3$N$_4$ grown on silicon and sapphire substrates in the same LPCVD furnace run, with a precursor gas ratio of NH$_3$:H$_2$SiCl$_2$~=~5:1 to ensure stoichiometric growth~\cite{Moille_OptLett_2021} at a temperature of 1048~K and a pressure of 300~Pa. In Fig.~\ref{Fig2}(a), we see that Si$_3$N$_4$ films on a silicon substrate remain crack-free for a thickness up to 380~nm, and exhibit tensile stress of 900~MPa to 1100~MPa. Above 400~nm, the films start to crack and stress measurements cannot be performed, with trench lines needed for successful growth for films in the 400~nm to 700~nm thickness range, tiling patterns needed for films above 700~nm (up to thicknesses near 1~$\mu$m), and a damascene process needed for even thicker films. While the data we have shown here is Si$_3$N$_4$ grown on 2.7~$\mu$m silicon dioxide on a silicon substrate, with the silicon dioxide formed by wet oxidation, the direct growth of Si$_3$N$_4$ on silicon shows similar tensile stress (to be expected given the silicon substrate is more than 100~$\times$ thicker than the silicon dioxide and hence its CTE dominates).   

In contrast to the above results on silicon, when using sapphire as the substrate, we are able to grow films up to 1960~nm without any cracks, as shown Fig.~\ref{Fig2}(a) (red). Importantly, the stress is measured to be $\approx$~200~MPa and shows no increase with thickness, suggesting that even thicker crack-free films can be grown. Moreover, we are able to grow thick Si$_3$N$_4$ films on intermediate layers of silicon dioxide and aluminum nitride on the sapphire substrate, confirming our understanding that the substrate CTE is the dominant consideration.

So far, the stress measurements of Si$_3$N$_4$ have conclusively shown that the sapphire substrate (with a much larger CTE than Si$_3$N$_4$) is advantageous in reducing tensile stress compared to the silicon substrate (with a close but relatively smaller CTE than Si$_3$N$_4$), as well as the fused silica substrate (smallest CTE of the three). While Fig.~\ref{Fig1}(c) seems to be a possible explanation, it is important to examine the stress data with temperature variation to corroborate this explanation. Due to technical constraints, we cannot readily measure the stress while the wafers are cooling down from 1023~K, but we can measure the stress when heating them up to 600~K, as shown in Figure~\ref{Fig2}(b), where the tool we use measures residual stress based on the curvature of the wafer. We find that the \xl{residual stress of the nitride film on silicon wafer} stays consistently high (at the $\approx$1000~MPa level), while \xl{the stress of the nitride film on the sapphire wafer} is much higher when heated, increasing from 200~MPa near room temperature to $\approx~$400~MPa at 500~K to 600~K. This significant change in stress for a sapphire substrate is indicative of its ability to relieve the Si$_3$N$_4$ film stress during cooling, as suggested in Fig.~\ref{Fig1}(c), and is consistent with the temperature-dependent CTE for the different materials as reported in the literature and shown in Fig.~\ref{Fig2}(c).


\begin{figure*}[t!]
\centering\includegraphics[width=0.96\linewidth]{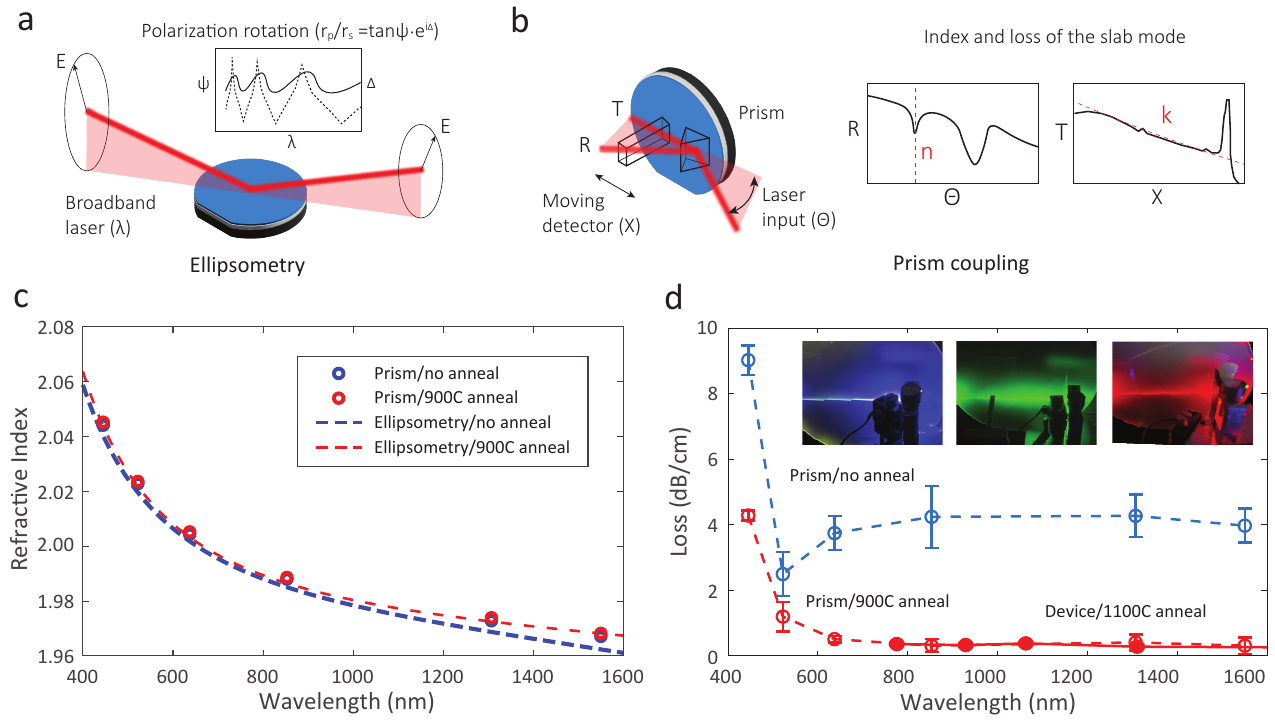}
\caption{\textbf{Ellipsometry and prism coupling measurement of SiNoSa wafers before and after annealing.} \textbf{a-b} Schematic illustrations of the two measurement approaches. \textbf{c,} Refractive index and loss of thin film characterized by ellipsometry (dashed lines) and prism coupling (open circles) before and after annealing. \textbf{d,} Measured loss determined via prism coupling (dashed lines) for thin films under different annealing conditions. \xl{The error bars represent two standard deviations} of the exponential fitting for the propagation loss of the fundamental slab modes. For comparison, the solid red circles \xl{and line} are for Si$_3$N$_4$ devices on a sapphire substrates, with more details provided in the subsequent figure. The inset images show the prism coupling measurements in the blue, green, and red spectral regions.}
\label{Fig5to4}
\end{figure*} 

\begin{figure*}[t!]
\centering\includegraphics[width=0.96\linewidth]{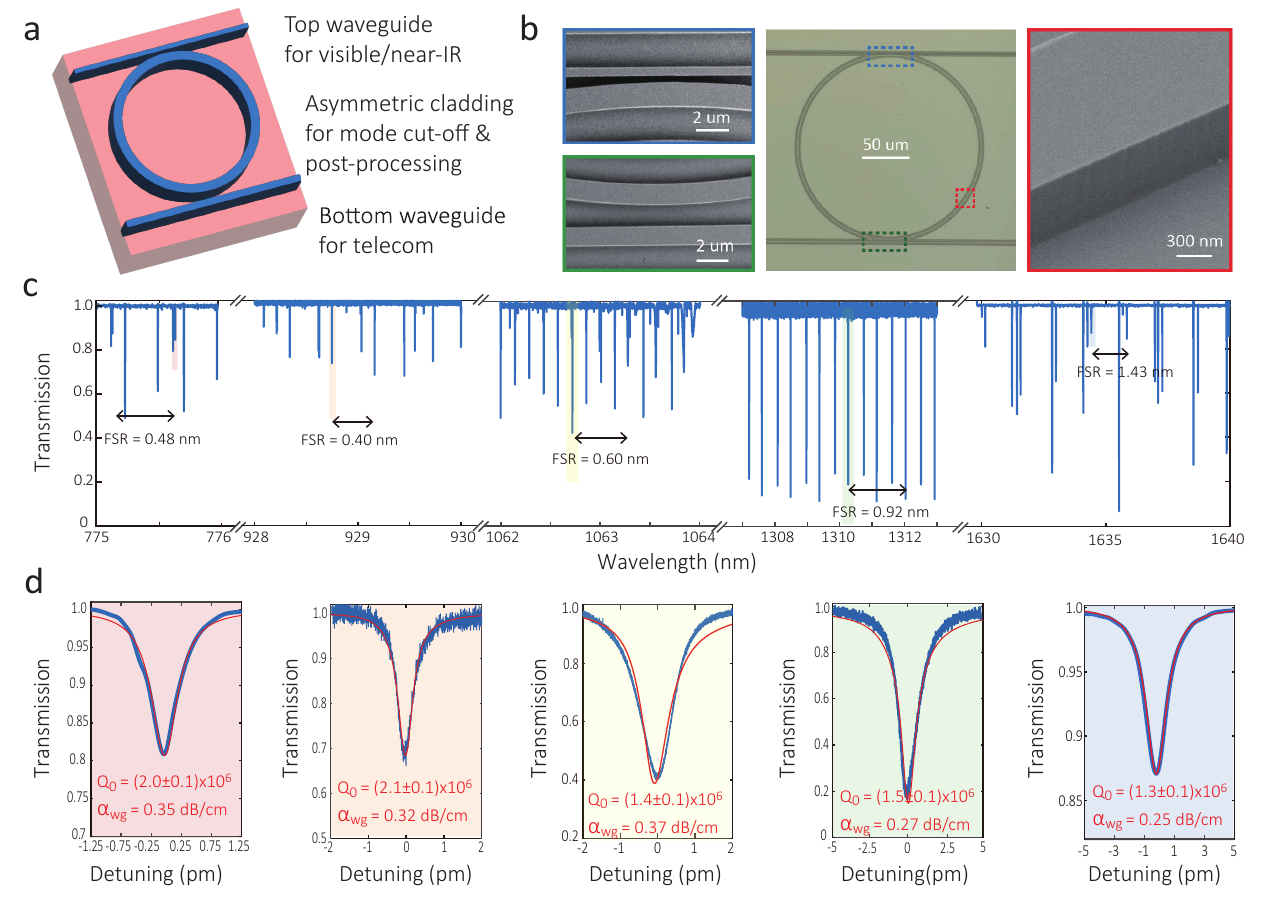}
\caption{\textbf{Fabricated photonic device in SiNoSa and its characterization.} \textbf{a,} Schematic illustration of a device. \textbf{b,} A microring (center, optical microscope image) with two waveguides coupling (left, scanning electron microscope images) from top and bottom with designs tailored to address different wavelength regions. The etched sidewall is straight and smooth (right scanning electron microscope image). This device has a ring radius of 150~$\mu$m, a ring width of 1.5~$\mu$m, and a thickness of 650~nm. \textbf{c,} The measured optical transmission traces show a good number of optical resonances with reasonable coupling using five tunable lasers around 775~nm, 930~nm, 1060~nm, 1310~nm, and 1630~nm. The 775~nm device has an outer radius of 100~$\mu$m, whereas all other devices have an outer radius of 150~$\mu$m. The fundamental transverse electric (TE) modes are identified based on a free-spectral range (FSR) analysis and comparison with simulations. \textbf{d,} Five fundamental transverse-electric modes in different bands are displayed and exhibit intrinsic optical quality factors above $1.3\times10^6$, corresponding to propagation losses between 0.25~dB/cm and 0.37~dB/cm. The uncertainties are \xl{one-standard deviation value} \xl{obtained from} the Lorentzian fitting the experimental data (light red lines in (d)).}
\label{Fig5}
\end{figure*}

Figure~\ref{Fig3} shows accompanying photographs and optical microscope images of Si$_3$N$_4$ pieces grown on different substrates to reinforce the stress analysis presented above. In fused silica, it is difficult to prevent crack formation even when trench lines are incorporated. As noted earlier, for silicon substrates, crack mitigation strategies such as trench lines and tiling patterns are needed to achieve Si$_3$N$_4$ thickness $>$~400~nm, and depending on the approach taken, cracking and film delamination remains a significant challenge beyond a certain thickness. In contrast, the Si$_3$N$_4$ grown on sapphire does not need trench lines at all, and show no cracks even close to diced edges.

\vspace{-0.15in}
\section{Photonic characterization}
\vspace{-0.15in}
\noindent We first characterize the LPCVD Si$_3$N$_4$ films on sapphire by both ellipsometry and prism coupling, as illustrated by Fig.~\ref{Fig5to4}(a,b). The refractive indices measured by both methods are found to be consistent with each other in Fig.~\ref{Fig5to4}(c). \xl{The film grown at 1048~K (775~$^{\circ}$C) has a wavelength-dependent refractive index of $n^2=1+2.886/[1-(0.13265/\bar{\lambda})^2]-0.02345\bar{\lambda}^2$, where $\bar{\lambda} = \lambda/(1~\mu \text{m})$. The film annealed at 1173~K (900~$^{\circ}$C) for three hours has a refractive index of $n^2=1+2.878/[1-(0.13705/\bar{\lambda})^2]-0.01116\bar{\lambda}^2$.}

Through the prism coupling measurements with a movable detector head, we find that the unannealed film shows optical losses above 2~dB/cm in general, and annealing reduces the optical losses at all wavelengths as shown in Fig.~\ref{Fig5to4}(d), to a value $<$1~dB/cm for wavelengths above 600~nm. The insets show the propagation of slab modes of a 660~nm SiNoSa film excited by blue, green, and red lasers. We note that it is challenging to characterize losses below 0.5~dB/cm in our prims coupling tool, although in principle a smaller loss can be extracted according to the tool specification.

Using knowledge of the Si$_3$N$_4$ refractive index and film propagation loss when grown on a sapphire substrate, we then fabricate Si$_3$N$_4$ photonic chips from a 1.8~cm $\times$ 2.2~cm sample diced from a 100~mm wafer, where the Si$_3$N$_4$ waveguiding layer is 650~nm thick and is grown on the sapphire substrate directly. The devices we fabricate are microring resonators (Fig.~\ref{Fig5}(a)), and Fig.~\ref{Fig5}(b) shows an optical microscope image of the device (center), along with scanning electron microscope (SEM) images of the top and bottom coupling waveguides (left) for visible/near-infrared and telecom wavelengths, respectively, and an SEM image of the etched sidewall (right). The sidewall smoothness is of a similar level to that observed for Si$_3$N$_4$ on thermally oxidized silicon substrates, with a good etch stop because sapphire is very resistant to dry etching using a fluorinated chemistry. The specific process we use is reactive ion etching with CHF$_3$ and O$_2$ gases at flow rates of 30~sccm and 2~sccm, respectively, and a power of 150~W at a pressure of 2~Pa (15~mTorr). The resulting etching rate is $<$~1~nm per minute for sapphire, $\approx$~30~nm per minute for Si$_3$N$_4$, and $>$~20~nm per minute for SiO$_2$. The selectivity ratio of etching Si$_3$N$_4$ with respect to ZEP520a (an electron beam lithography resist) and with respect to DUV210 (a deep ultra-violet photoresist) is around 1.3:1 and 1.5:1, respectively~\cite{NIST_disclaimer}.

We measure the cavity transmission through swept-wavelength spectroscopy with tunable external cavity diode lasers at different colors, where light is coupled to the chips with lensed optical fibers (Fig.~\ref{Fig5}(c)-(d)). At 775~nm, we find the fundamental transverse-electric mode has an intrinsic optical quality factor of (2.0$\pm$0.1)$\times$10$^6$ in a microring with a radius of 100~$\mu$m, a ring width of 1625~nm, and a free spectral range of 0.48~nm. The propagation loss is thus calculated to be (0.35$\pm$0.02)~dB/cm at 775~nm. The intrinsic quality factor measured at 929~nm is (2.1$\pm$0.1)$\times$10$^6$, with a corresponding propagation loss of (0.32$\pm$0.02)~dB/cm. 
The intrinsic quality factor measured at 1062~nm is (1.4$\pm$0.1)$\times$10$^6$, with a corresponding propagation loss of (0.37$\pm$0.03)~dB/cm. The quality factor near 1310~nm (O-band) is (1.5$\pm$0.1)$\times$10$^6$ (propagation loss is (0.27$\pm$0.02) dB/cm). The highest intrinsic quality factor near 1630~nm (the U-band) is (1.3$\pm$0.1)$\times$10$^6$, which corresponds to (0.25$\pm$0.02)~dB/cm. This value is similar to our previous work using similar recipe, but with Si$_3$N$_4$ on an oxidized silicon wafer. Data for the devices fabricated on a sapphire substrate are also shown in Fig.~\ref{Fig5to4}(d) for comparison against the thin film loss, while Fig.~\ref{Fig1}(a) contains the above loss per unit length values as a function of wavelength and in comparison to other results from SiN (not necessarily stoichiometric) films.

We note that \xl{high confinement and low confinement Si$_3$N$_4$ devices can have even lower loss values} below 0.01~dB/cm, by either using ultra-thick Si$_3$N$_4$ (typically with chemical mechanical polishing to avoid sidewall scattering), or using ultra-thin Si$_3$N$_4$ where the mode is mainly in the high quality silicon dioxide cladding rather than the Si$_3$N$_4$~\cite{Ji_APLPhoton_2021}. While both thick and thin silicon nitride can achieve ultra-low loss, and thin nitride is particularly useful in processes such as stimulated Brillouin scattering~\cite{Puckett_NatCommun_2021} and self-injection locking~\cite{Li_Optica_2023}, most linear optical circuits~\cite{Xiang_Optica_2020, Churaev_NatCommun_2023, Xanadu_Nature_2025, PsiQuantum_Nature_2025} and wide-band nonlinear optical processes~\cite{Gaeta_NatPhoton_2019, Chang_NatPhoton_2022,Lu_Optica_2019, Domeneguetti_Optica_2021,Lu_NatPhoton_2021, Hu_SciAdv_2022}, including Kerr frequency comb generation, optical parametric oscillation, and harmonic generation, require moderately confined Si$_3$N$_4$ for a balance of dispersion engineering, coupling design, and optical quality factor. We therefore have focused on moderately-confined Si$_3$N$_4$ in this work, while future studies could explore loss limits in the other confinement regimes.


\vspace{-0.15in}
\section{Conclusions}
\noindent We have demonstrated crack-free stoichiometric silicon nitride films up to 2~$\mu$m thickness and without any stress mitigation approaches through LPCVD growth on sapphire substrates. Stress measurements for different film thickness and temperature-dependent measurements of stress for a fixed thickness support the above observation and point to the higher CTE substrate that sapphire provides as the key enabler for thick Si$_3$N$_4$ growth. We further show its advantages in photonics fabrication, demonstrating low optical losses on par with current Si$_3$N$_4$ on silicon devices within the same confinement regime across a broad range of wavelengths between 780~nm and 1650~nm. From these measurements, we conclude that Si$_3$N$_4$ on sapphire is an alternative wide-band photonics platform, with benefits including wavelength accessibility in the deep visible and mid-infrared, elimination of the broadband Raman noise of the lower silicon dioxide cladding for quantum applications, and wafer-level fabrication of thick films without stress mitigation. 

\section{Acknowledgments}
X.L. acknowledges Richard Kasica for his help with furnace operation, and Joseph Wikar for his help with wafer dicing. The authors acknowledge Yang Xu and Jordan Stone for helpful discussions. The authors acknowledge the NIST-on-a-chip program and AFRL through award FA9453-20-2-0001, sub-award 2RQM8. X.L. acknowledges the funding support from Maryland Industrial Partnerships (MIPS). This work was performed at the NIST Center for Nanoscale Science and Technology (CNST), and in part at the Cornell NanoScale Facility (CNF), funded by the National Science Foundation (NNCI-2025233).






\bibliographystyle{osajnl}
\bibliography{SiNOS}

\end{document}